\documentclass[apjl]{emulateapj}
\usepackage{natbib}
\usepackage{mathptmx}

\usepackage{graphicx}

\def\h2o{\ifmmode x_{\rm H_2O} \else $x_{\rm H_2O}$ \fi}
\def\iuv{\ifmmode I_{\rm UV} \else $I_{\rm UV}$ \fi}
\def\in{\ifmmode I_{\rm UV}/n_4 \else $I_{\rm UV}/n_4$ \fi}
\def\mw{\ifmmode x_{\rm H_2O, MW} \else $x_{\rm H_2O, MW}$ \fi}
\newcommand\hh{\ifmmode {\rm H_2} \else H$_2$ \fi}
\def\no{\ifmmode {N_{\rm HI}} \else $N_{\rm HI}$ \fi}
\def\nt{\ifmmode {N_{\rm H_2}} \else $N_{\rm HI}$ \fi}
\def\so{\ifmmode {\Sigma_{\rm HI}} \else $\Sigma_{\rm HI}$ \fi}
\def\st{\ifmmode {\Sigma_{\rm H_2}} \else $\Sigma_{\rm H_2}$ \fi}
\renewcommand\ss{\ifmmode {\Sigma_{\rm tot}} \else $\Sigma_{\rm tot}$ \fi}
\def\msun{\ifmmode {\rm M_{\odot}}\else $\rm M_{\odot}$\fi}
\def\mpc{\ifmmode {\rm M_{\odot}/pc^2}\else $\rm M_{\odot}/pc^2$\fi}
\def\tra{\ifmmode {\rm HI-to-H_2}\else H{\small I}-to-H$_2$ \fi}

\begin{document}

\title{Water Formation During the Epoch of First Metal Enrichment}
\author{
Shmuel Bialy$^\star$\altaffilmark{1}, Amiel Sternberg\altaffilmark{1} and Abraham Loeb\altaffilmark{1,2} 
}

\altaffiltext{1}
{Raymond and Beverly Sackler School of Physics \& Astronomy,
Tel Aviv University, Ramat Aviv 69978, Israel}
\altaffiltext{2}
{Astronomy Department, Harvard University, 60 Garden St., Cambridge, MA 02138, USA}

\email{$^\star$shmuelbi@mail.tau.ac.il}

\slugcomment{ApJ Letters accepted}

\begin{abstract}
We demonstrate that high abundances of water vapor could have existed in extremely low metallicity (10$^{-3}$ solar) partially shielded gas, during the epoch of first metal enrichment of the interstellar medium of galaxies at high redshifts.
\end{abstract}


\section{Introduction}
Water is an essential ingredient for life as we know it \citep{Kasting2010}. 
In the interstellar medium (ISM) of the Milky-Way and also in external galaxies, water has been observed in the gas phase and as grain surface ice in a wide variety of environments. 
These environments include diffuse and dense molecular clouds, photon-dominated regions (PDRs), shocked gas, protostellar envelopes, and disks (see review by \citealt{vanDishoeck2013}).

In diffuse and translucent clouds, H$_2$O is formed mainly in gas phase reactions via ion-molecule sequences \citep{Herbst1973}. 
The ion-molecule reaction network is driven by cosmic-ray or X-ray ionization of H and H$_2$, that leads to the formation of  H$^+$ and H$_3^+$ ions. These interact with atomic oxygen and form OH$^+$. A series of abstractions then lead to the formation of H$_3$O$^+$, which makes OH and H$_2$O through dissociative recombination.   
This formation mechanism is generally not very efficient, and only a small fraction of the oxygen is converted into water, 
the rest remains in atomic form, or freezes out as water ice \citep{Hollenbach2009}.

\citet{Sonnentrucker2010} showed that the abundance of water vapor within diffuse clouds in the Galaxy is remarkably constant, with $x_{\rm H_2O} \sim 10^{-8}$, that is $\sim 0.1 \%$ of the available oxygen. Here $x_{\rm H_2O}$ is the H$_2$O number density relative to the total hydrogen nuclei number density.
Towards the Galactic center this value can be enhanced by up to a factor of $\sim 3$ \citep{Monje2011, Sonnentrucker2013}. 

At temperatures $\gtrsim 300$~K, H$_2$O may form directly via the neutral-neutral reactions, O + H$_2$ $\rightarrow$ OH + H, followed by OH + H$_2$ $\rightarrow$ H$_2$O + H. 
This formation route is particularly important in shocks, where the gas heats up to high temperatures, and can drive most of the oxygen into H$_2$O \citep{Draine1983, Kaufman1996}.

Temperatures of a few hundreds K are also expected in very low metallicity gas environments, with elemental oxygen and carbon abundances of $\lesssim 10^{-3}$ solar \citep{Bromm2003, Omukai2005, Glover2014}, associated with the epochs of the first enrichment of the ISM with heavy elements, in the first generation of galaxies at high redshifts \citep{Loeb2013}. 
At such low metallicities, cooling by fine structure transitions of metal species such as the [CII] $158$~$\mu$m line, and by rotational transitions of heavy molecules such as CO, becomes inefficient and the gas remains warm.

Could the enhanced rate of H$_2$O formation via the neutral-neutral sequence in such warm gas, compensate for the low oxygen abundance at low metallicities? 

\citet{Omukai2005} studied the thermal and chemical evolution of collapsing gas clumps at low metallicities.
They found that for models with gas metallicities of $10^{-3}-10^{-4}$ solar, $x_{\rm H_2O}$ may reach $10^{-8}$, but only if the density, $n$, of the gas approaches $10^{8}$~cm$^{-3}$. 
Photodissociation of molecules by far-ultraviolet (FUV) radiation was not included in their study. 
While at solar metallicity dust-grains shield the interior of gas clouds from the FUV radiation, at low metallicities photodissociation by FUV becomes a major removal process for H$_2$O. 
H$_2$O photodissociation produces OH, which is then itself photodissociated into atomic oxygen.

\citet{Hollenbach2012} developed a theoretical model to study the abundances of various molecules, including H$_2$O, in PDRs. 
Their model included many important physical processes, such as freezeout of gas species, grain surface chemistry, and also photodissociation by FUV photons. 
However they focused on solar metallicity.
Intriguingly, \citet{Bayet2009} report a water abundance close
to $10^{-8}$ in the optically thick core of their single PDR model for 
a low metallicity of 10$^{-2}$ (with $n=10^3$~cm$^{-3}$).  
However, Bayet et al.~did not investigate the effects of temperature and UV intensity variations on the water abundance in the low metallicity regime.

Recently, a comprehensive study of molecular abundances for the bulk ISM gas as functions of the metallicity, were studied by \citet{Penteado2014} and \citet[][hereafter BS15]{Bialy2014}, 
these models however, focused on the ``low temperature" ion-molecule formation chemistry.

In this {\it Letter} we present results for the H$_2$O abundance in low metallicity gas environments, for varying temperatures, FUV intensities and gas densities. 
We demonstrate the importance of the onset of the neutral-neutral formation sequence, and explore the role of the FUV field.
We find that for temperatures $T$ in the range $250-350$ K, H$_2$O may be abundant, comparable to or higher than that found in diffuse Galactic clouds, provided that the FUV intensity to density ratio is smaller than a critical value.
In \S~\ref{sec:method} we discuss our physical and chemical model. We present our results in \S~\ref{sec:Results}. Finally we summarize and discuss our conclusions in \S~\ref{sec:summary}.


%


\section{Model Ingredients}
\label{sec:method}
We calculate the abundance of gas-phase H$_2$O for low metallicity gas parcels, that are exposed to external FUV radiation and cosmic-ray and/or X-ray fluxes. Given our chemical network we solve the steady state rate equations using our dedicated Newton-based solver, and obtain $x_{\rm H_2O}$ as function of $T$ and the FUV intensity to density ratio.

We adopt a 10$^5$~K diluted blackbody spectrum, representative of radiation produced by massive Pop-III stars.
The photon density in the 6-13.6 eV interval, is $n_{\gamma} \equiv n_{\gamma, 0}\iuv$, where $n_{\gamma, 0} = 6.5 \times 10^{-3}$~photons~cm$^{-3}$
is the photon density in the interstellar radiation field \citep[ISRF, ][]{Draine2011}, and \iuv is the ``normalized intensity". Thus $\iuv=1$ corresponds to the FUV intensity in the Draine ISRF.

Cosmic-ray and/or X-ray ionization drive the ion-molecule chemical network. 
We assume an ionization rate per hydrogen nucleon $\zeta$ (s$^{-1}$).
In the Galaxy, \citet{Dalgarno2006} and \citet{Indriolo2012} showed that $\zeta$ lies within the relatively narrow range $10^{-15}-10^{-16}$~s$^{-1}$. We therefore introduce the ``normalized ionization rate" $\zeta_{-16} \equiv (\zeta/10^{-16}$~s$^{-1})$.
The ionization rate and the FUV intensity are likely correlated, as both should scale with the formation rate of massive OB stars. We thus set $\zeta_{-16} = \iuv$ as our fiducial case but also consider the cases $\zeta_{-16} = 10^{-1} \iuv$ and $\zeta_{-16} = 10 \iuv$.

Dust shielding against the FUV radiation becomes ineffective at low metallicities. However, self absorption in the H$_2$ lines may significantly attenuate the destructive Lyman Werner (11.2-13.6 eV) radiation \citep{Draine1996, Sternberg2014} and high abundances of H$_2$ may be maintained even at low metallicity (BS15).  
In the models presented here we assume an H$_2$ shielding column of at least $5 \times 10^{21}$~cm$^{-2}$. 
(For such conditions CO is also shielded by the H$_2$.)
The LW flux is then attenuated by a self-shielding 
factor of $f_{shield}\sim 10^{-8}$ and the H$_2$ photodissociation rate is only $5.8 \times 10^{-19}  I_{\rm UV}$~s$^{-1}$.  With this assumption
H$_2$ photodissociation by LW photons is negligible compared to cosmic-ray and/or X-ray ionization as long as $\iuv < 85 \zeta_{-16}$.

However, even when the Lyman Werner band is fully blocked, OH and H$_2$O are photodissociated because their energy thresholds for photodissociation are $6.4$ and 6 eV, respectively.
For the low metallicities that we consider, photodissociation is generally the dominant removal mechanism for H$_2$O and OH.
We adopt the calculated OH and H$_2$O photodissociation rates calculated by BS15. 

We assume thermal and chemical steady states. 
In the Milky Way, the bulk of the ISM gas is considered to be at approximate thermal equilibrium, set by cooling and heating processes. 
We discuss the relevant chemical and thermal time-scales in \S \ref{sub:time scales}.

Given the above mentioned assumptions, the steady state solutions for the species abundances depend on only two parameters, the temperature $T$ and the intensity to density ratio $\in$. Here $n_4 \equiv (n/10^4$~cm$^{-3})$ is the total number density of hydrogen nuclei normalized to typical molecular cloud densities.
$T$ and \in form our basic parameter space in our study.

\subsection{Chemical Model}
\label{sub:chemical model}

We consider a chemical network of 503 two-body reactions, between 56 atomic, molecular and ionic species of H, He, C, O, S, and Si. We assume cosmological elemental helium abundance of 0.1 relative to hydrogen (by number).
For the metal elemental abundances we adopt the \citet{Asplund2009} photospheric solar abundances, multiplied by a metallicity factor $Z'$ (i.e., $Z'=1$ is solar metallicity). 
In our fiducial model we assume $Z'=10^{-3}$, but we also explore cases with $Z'=10^{-2}$ and $Z'=10^{-4}$.
Since our focus here is on the very low metallicity regime, where dust grains play a lesser role, 
we neglect any depletion on dust grains, and dust-grain chemistry (except for H$_2$, as discussed further below). 
Direct and induced ionizations and dissociations by the cosmic-ray / X-ray field ($\propto \zeta$) are included.   
For the gas phase reactions, we adopt the rate coefficients given by the UMIST 2012 database \citep{McElroy2013}.

The formation of heavy molecules relies on molecular hydrogen.
We consider two scenarios for H$_2$ formation: (a) pure gas-phase formation; and (b) gas-phase formation plus formation on dust grains.
In gas-phase, radiative-attachment
\begin{equation}
	\label{R: H- form}
	{\rm H  \; + \;  e  \; \rightarrow \;  H^- \;  + \;  \nu}	
\end{equation}
followed by associative-detachment
\begin{equation}
	\label{R: H2 form gas}
	{\rm H^- \;  + \;  H \;  \rightarrow \;  H_2  \; + \; e} \, 
\end{equation} is the dominant H$_2$ formation route.

H$_2$ formation on the surface of dust grains is considered to be the dominant formation channel in the Milky way.
We adopt a standard rate coefficient 
\begin{equation}
	\label{eq:R_0}
	R \ \simeq \ 3 \times 10^{-17} \ T_2^{1/2} \ Z' \ {\rm cm^3 \ s^{-1}}
\end{equation} \citep{Hollenbach1971, Jura1974, Cazaux2002}, where $T_2 \equiv (T/100$~K$)$.
 In this expression we assume that the 
 dust-to-gas ratio is linearly proportional to the 
 metallicity $Z'$.  Thus,
in scenario (b) H$_2$ formation on dust grains dominates even for $Z'=10^{-3}$.
Scenario (a) is the limit where the gas-phase channel dominates, as appropriate for dust-free environments or for superlinear dependence of the dust-to-gas ratio on $Z'$ (see also BS15).

\subsection{Time scales}
\label{sub:time scales}
The timescale for the system to achieve chemical steady state is dictated by the relatively long H$_2$ formation timescale.
In the gas phase (scenario [a]) it is
\begin{equation}
	\label{eq: time scale gas}
	t_{\rm H_2} \, = \, \frac{1}{k_2 \, n \, x_{\rm e}} \, \approx \, 8 \times 10^8 \; \zeta_{-16}^{-1/2} \ n_4^{-1/2} \, T_2^{-1} \ \rm{yr} \ ,
\end{equation} 
where $x_{\rm e}=2.4 \times 10 ^{-5} (\zeta_{-16}/n_4)^{1/2} T_2^{0.38}$ is the electron fraction as set by ionization recombination equilibrium, and $k_2 \approx 1.5 \times 10^{-16} T_2^{0.67}$~cm$^3$~s$^{-1}$ is the rate coefficient for reaction (\ref{R: H- form}). 

For formation on dust grains ($\propto Z'$), the timescale is generally shorter, with
\begin{equation}
	\label{eq: time scale grain}
	t_{\rm H_2} \, = \, \frac{1}{R \, n} \, \approx \, 10^8 \; T_2^{-1/2} \ n_4^{-1} \, \left( \frac{10^{-3}}{Z'} \right) \ \rm{yr} \ .
\end{equation}
Gas clouds with lifetimes $t \gg t_{\rm H_2}$ will reach chemical steady state.

\begin{figure*}
	\centering
	\includegraphics[width=1\textwidth]{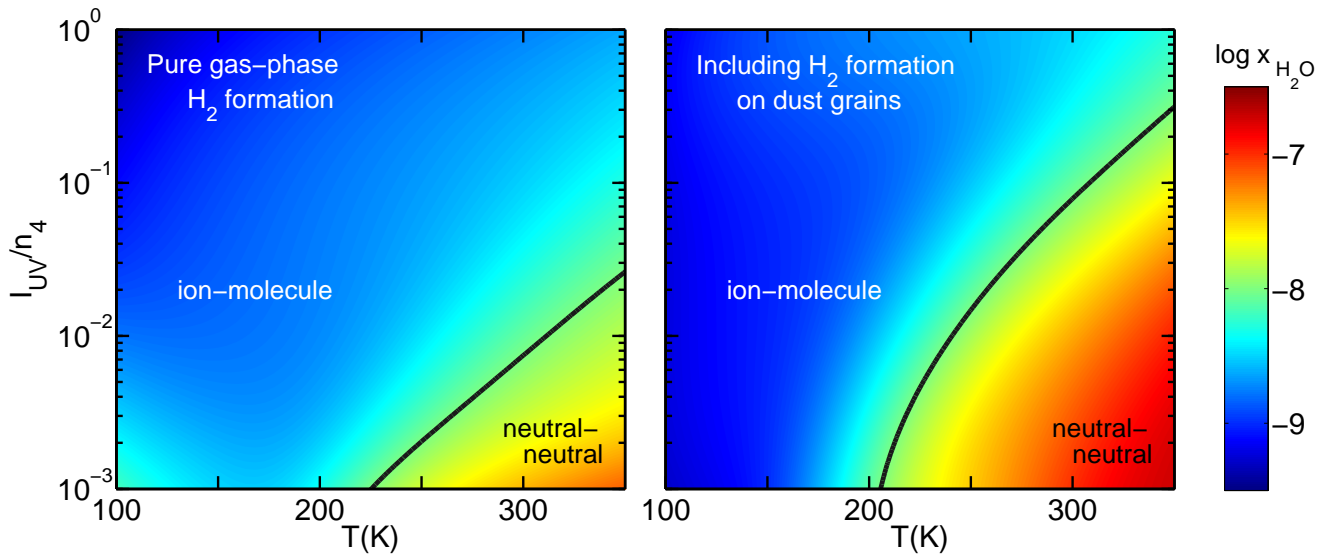}
	\caption{The fractional H$_2$O abundance $\h2o$ as a function of $T$ and $\in$, for $Z'=10^{-3}$ and $\zeta_{-16}/\iuv =1$, assuming pure gas-phase chemistry (scenario [a] - left panel), and including H$_2$ formation on dust grains (scenario [b] - right panels).
		In both panels, the solid line indicates the 10$^{-8}$ contour, which is a characteristic value for the H$_2$O gas phase abundance in diffuse clouds in the Milky-Way.
		At high temperatures (or low \in values) the neutral-neutral reactions become effective and \h2o rises.
	}
	\label{fig: H2O_T_UV}
\end{figure*} 

The relevant timescale for thermal equilibrium is the cooling timescale. 
For low metallicity gas with $Z'=10^{-3}$, the cooling proceeds mainly via H$_2$ rotational excitations \citep{Glover2014}.
If the cooling rate per H$_2$ molecule (in erg s$^{-1}$) is $W(n,T)$,
then the cooling timescale is given by    
\begin{equation}
	\label{eq: time scale H2 cooling}
	t_{\rm cool} \ = \ \frac{k_B \ T}{W(n,T)} \ .
\end{equation} Here $k_B$ is the Boltzmann constant.
	For $n=10^4$~cm$^{-3}$, and $T=300$~K, $W \approx 5 \times 10^{-25} (x_{\rm H_2}/0.1)$~erg~s$^{-1}$ \citep{LeBourlot1999}, and the cooling time is very short $\approx 2 \times 10^3 (0.1/x_{\rm H_2})$ yr. For densities much smaller than the critical density for H$_2$ cooling, $W \propto n$ and $t_{\rm cool} \propto 1/n$. In the opposite limit, $W$ saturates and $t_{\rm cool}$ becomes independent of density. We see that generally $t_{\rm cool} \ll t_{\rm H_2}$.
	
	Because the free fall time
	\begin{equation}
		t_{ff} \ = \ \left( \frac{3 \pi}{32 G \rho} \right)^{1/2} \ = \ 5 \times 10^5 \ n_4^{-1/2} \  {\rm yr} \ ,
	\end{equation}
	is generally much shorter than $t_{\rm H_2}$, chemical steady state may be achieved only in stable, non-collapsing clouds, with lifetimes $\gg t_{ff}$. 
	Obviously both $t_{\rm H_2}$ and $t_{\rm cool}$ must be also shorter than  
	the Hubble time at the redshift of interest.

\section{Results}
\label{sec:Results}

\begin{figure*}
	\centering
	\includegraphics[width=0.8\textwidth]{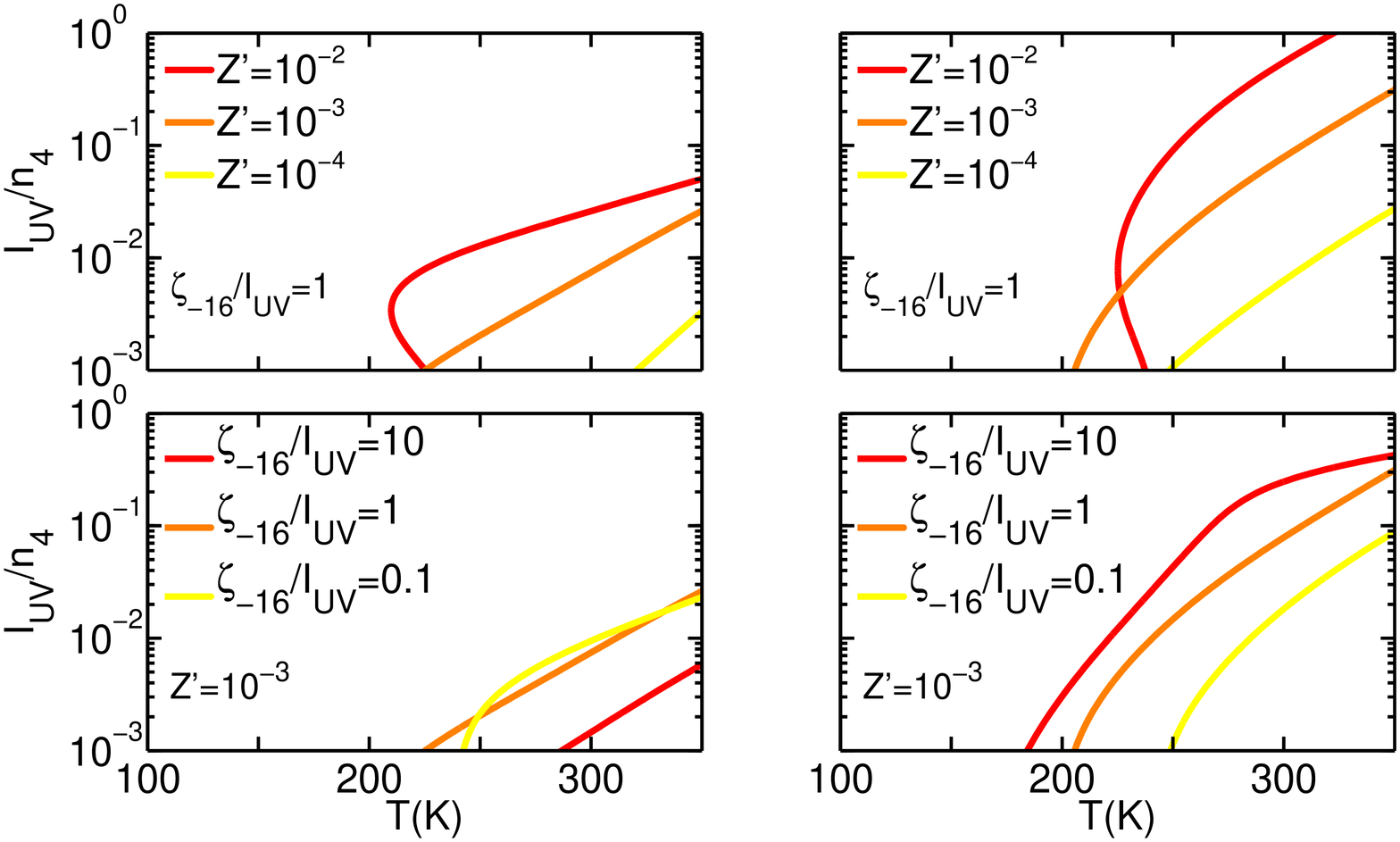}
	\caption{The $\h2o=10^{-8}$ contour, for variations in $Z'$ (upper panels) and in $\zeta_{-16}/\iuv$ (lower panels), assuming pure gas-phase chemistry (scenario [a] - left panels), and including H$_2$ formation on dust grains (scenario [b] - right panels).}
	\label{fig: variations}
\end{figure*}

Next we present and discuss our results for the steady state, gas-phase H$_2$O fraction $x_{\rm H_2O} \equiv n_{\rm H_2O}/n$, as function of temperature $T$, and the FUV intensity to density ratio $\in$.

\subsection{\h2o as a function of $T$ and \in}
Figure \ref{fig: H2O_T_UV} shows $\log_{10}(x_{\rm H_2O})$ contours for the two scenarios described in \S~\ref{sub:chemical model}. In one H$_2$ forms in pure gas-phase (scenario [a] - left panel), and in the other H$_2$ forms also on dust-grains (scenario [b] - right panel). Our fiducial parameters are $Z'=10^{-3}$ and $\zeta_{-16}=\iuv$.
At the upper-left region of the parameter space, $\h2o$ is generally low $\lesssim 10^{-9}$. 
In this regime, H$_2$O forms through the ion-molecule sequence, that is operative at low temperatures.  
In the lower right corner, the neutral-neutral reactions become effective and \h2o rises.

In both panels, the solid line highlights the $\h2o=10^{-8}$ contour, which resembles the H$_2$O gas phase abundance in diffuse and translucent Galactic clouds. This line also delineates the borderline between the regimes where H$_2$O forms via the ``cold" ion-molecule sequence, and the ``warm" neutral-neutral sequence.
The temperature range at which the neutral-neutral sequence kicks-in is relatively narrow $\sim 250-350$~K, because the neutral-neutral reactions are limited by energy barriers that introduce an exponential dependence on temperature.

The dependence on \in is introduced because the FUV photons photodissociate OH and H$_2$O molecules and therefore increase the removal rate of H$_2$O and at the same time suppress formation via the OH + H$_2$ reaction. This gives rise to a critical value for $\in$ below which H$_2$O may become abundant. 
    
For pure gas-phase H$_2$ formation (scenario [a] - left panel), the gas remains predominantly atomic, and H$_2$O formation is less efficient. In this case $\h2o \gtrsim 10^{-8}$ only when \in is smaller than a critical value of
\begin{equation}
\left( I_{\rm UV}/n_4 \right)_{\rm crit}^{\rm (a)} \ = \ 2 \times 10^{-2} \ \ \ .
\end{equation}
However, when H$_2$ formation on dust is included (scenario [b] - right panel), the hydrogen becomes fully molecular, and H$_2$O formation is then more efficient.
In this case $\h2o$ may reach 10$^{-8}$ for $\in$ smaller than  
\begin{equation}
\left( I_{\rm UV}/n_4 \right)_{\rm crit}^{\rm (b)} \ = \ 3 \times 10^{-1} \ \ \ ,
\end{equation}
an order of magnitude larger than for the pure-gas phase formation scenario.

%
%
%

\subsection{Variations in $Z'$ and $\zeta_{-16}/\iuv$}
In Figure \ref{fig: variations} we investigate the effects of variations in the value of $Z'$ and the normalization $\zeta_{-16}/\iuv$.
The Figure shows the $\h2o=10^{-8}$ contours 
for scenarios (a) (left panels) and (b) (right panels).
As discussed above, H$_2$O is generally more abundant in scenario (b) because the hydrogen is fully molecular in this case, and therefore the 10$^{-8}$ contours are located at higher \in values in both right panels. 

The upper panels show the effect of variations in the metallicity value $Z'$, for our fiducial normalization $\zeta_{-16}=\iuv$.
In both panels, the oxygen abundance rises, and $\h2o$ increases with increasing $Z'$. 
Thus at higher $Z'$, the 10$^{-8}$ contours shift to lower $T$ and higher $\in$ and vice versa.
An exception is the behavior of the $Z'=10^{-2}$ curve, for which the metallicity is already high enough so that reactions with metal species dominate H$_2$O removal for $\in \lesssim 10^{-2}$.
The increase in metallicity then results in a {\it decrease} of the H$_2$O abundance, and the 10$^{-8}$ contour shifts to the right. 
For $\in \gtrsim 10^{-2}$ removal by FUV dominates and the behavior is similar to that in the $Z'=10^{-3}$ and $Z'=10^{-4}$ cases.

The lower panels show the effects of variations in the ionization rate normalization $\zeta_{-16}/\iuv$ for our fiducial metallicity value of $Z'=10^{-3}$. 
First we consider the pure gas phase formation case (scenario [a] - lower left panel). 
For the two cases $\zeta_{-16}/\iuv = 1$ and 10$^{-1}$, the H$_2$O removal is dominated by FUV photodissociation and therefore is independent of $\zeta$. 
As shown by BS15, the H$_2$O formation rate is also independent of $\zeta$ when the H$_2$ forms in the gas-phase. Therefore \h2o is essentially independent of $\zeta$ and the contours overlap.  
For the high ionization rate $\zeta_{-16}/\iuv = 10$, the proton abundance becomes high, and H$_2$O reactions with H$^+$ dominate H$_2$O removal.
In this limit \h2o decreases with $\zeta$ and the 10$^{-8}$ contour moves down.

When H$_2$ forms on dust (scenario [b] - lower right panel), the H$_2$O formation rate via the ion-molecule sequence is proportional to the H$^+$ and H$_3^+$ abundances, which rise with $\zeta$. 
Since the gas is molecular, the proton fraction is low and the removal is always dominated by FUV photodissociations (independent of $\zeta$). 
Therefore, in this case $\h2o$ {\it increases} with $\zeta_{-16}/\iuv$ and the curves shift up and to the left, toward lower $T$ and higher $\in$. 

\section{Summary and Discussion}
\label{sec:summary}
In this {\it Letter} we have demonstrated that the H$_2$O gas phase abundance may remain high even at very low metallicities of $Z' \sim 10^{-3}$. The onset of the efficient neutral-neutral formation sequence at $T \sim 300$~K, may compensate for the low metallicity, and form H$_2$O in abundance similar to that found in diffuse clouds in the Milky Way. 

We have considered two scenarios for H$_2$ formation, representing two limiting cases, one in which H$_2$ is formed in pure gas phase (scenario [a]), and one in which H$_2$ forms both in gas-phase and on dust grains, assuming that the dust abundance scales linearly with $Z'$ (scenario [b]).
Recent studies by \citet{Galametz2011}, \citet{Herrera-Camus2012} and \citet{Fisher2014}, suggest that the dust abundance might decrease faster then linearly with decreasing $Z'$. 
As shown by BS15, for $Z'=10^{-3}$ and dust abundance that scales as $Z'^{\beta}$ with $\beta \geq 2$, H$_2$ formation is dominated by the gas phase formation channel.
Therefore our scenario (a) is also applicable for models in which dust grains are present, with an abundance that scales superlinearly with $Z'$.
For both scenarios (a) and (b) we have found that the neutral-neutral formation channel yields $\h2o\gtrsim 10^{-8}$, provided that $\in$ is smaller than a critical value. 
For the first scenario we have found that this critical value is $(I_{\rm UV}/n_4)_{\rm crit} = 2 \times 10^{2}$. For the second scenario $(I_{\rm UV}/n_4)_{\rm crit} = 3  \times 10^{-1}$.

In our analysis we have assumed that the system had reached a chemical steady state. 
For initially atomic (or ionized) gas, this assumption 
offers the best conditions for the formation of molecules. 
However, chemical steady state might not always be achieved within a cloud lifetime or even within the Hubble time. 
The timescale to achieve chemical steady state (from an initially 
dissociated state)
is dictated by the H$_2$ formation process, and is generally long at low metallicities. 
For $Z'=10^{-3}$ and our fiducial parameters, the timescales for both scenarios are of order of a few 10$^{8}$ years (see e.g.~\citealt{Bell2006}), and are comparable to the age of the Universe at redshifts of $\sim 10$. 
The generically high water abundances we find for warm
conditions and low metallicities will be maintained in
dynamically evolving systems so long as they remain
H$_2$-shielded.


Our results might have interesting implications for the question of how early could have life originated in the Universe \citep{Loeb2014}. 
Our study addresses the first step of H$_2$O formation in early enriched, molecular gas clouds. 
If such a cloud is to collapse and form a protoplanetary disk, some of the H$_2$O may make its way to the surfaces of forming planets \citep{vanDishoeck2014}.
However, the question of to what extent the H$_2$O molecules that were formed in the initial molecular clouds, are preserved all the way through the process of planet formation, is beyond the scope of this paper.

\acknowledgments
S.B.~acknowledges support from the Raymond and Beverly Sackler Tel Aviv University -- Harvard/ITC Astronomy Program.
A.L. thanks the Sackler professorship at Tel Aviv University for generous support, as well as the NSF grant AST-1312034.
This work was also supported in part by the DFG via German -- Israeli 
Project Cooperation grant STE1869/1-1/GE625/15-1, and by a
PBC Israel Science Foundation I-CORE Program grant 1829/12.

\end{document}